\documentclass[
prc,%
10pt,%
final,%
notitlepage,%
oneside,%
twocolumn,%
nobibnotes,%
nofootinbib,
superscriptaddress,%
floatfix,%
showkeys,%
showpacs]%
{revtex4}
\usepackage{color}
\usepackage{amsfonts}
\usepackage{amsbsy}
\usepackage{mathrsfs}
\usepackage{graphicx}
\def\lsim{\mathrel{\rlap{
\lower4pt\hbox{\hskip-3pt$\sim$}}
    \raise1pt\hbox{$<$}}}     
\def\gsim{\mathrel{\rlap{
\lower4pt\hbox{\hskip-3pt$\sim$}}
    \raise1pt\hbox{$>$}}}     

\begin{document}
\title{Baryon Stopping as a Probe of Deconfinement Onset in Relativistic Heavy-Ion Collisions}%
\author{Yu.B. Ivanov}\thanks{e-mail: Y.Ivanov@gsi.de}
\affiliation{Kurchatov Institute, 
Moscow RU-123182, Russia}
\begin{abstract}
It is argued that an irregularity in the baryon stopping is a natural consequence of 
{
onset of deconfinement 
}
occurring in the compression stage of a nuclear collision. It is a combined effect 
of the softest point inherent in an equation of state (EoS) with 
{
a deconfinement transition 
}
and a change in the nonequilibrium dynamics from hadronic to partonic transport.  
Thus, this irregularity 
is a signal from a hot and dense stage of the nuclear collision.  
In order to illustrate this proposition, calculations within the three-fluid model were performed 
with three different EoS's: a purely hadronic EoS, an EoS with a first-order phase transition  
and that with a smooth crossover transition.
It is found that predictions within the  first-order-transition scenario
indeed reveal an  a strong irregularity in the 
incident energy dependence of the form of the net-proton rapidity distributions in central collisions. 
This behavior is in contrast to that for the hadronic scenario, 
where the distribution form gradually evolve, displaying no irregularity. 
The case of the crossover EoS is intermediate. Only  
a weak irregularity  takes place. 
Experimental data also exhibit a trend of similar irregularity, 
which is however based on still preliminary data at energies of 20$A$ GeV and 30$A$ GeV.
\end{abstract}
\maketitle

\section{Introduction}

Onset of deconfinement in relativistic heavy-oin collisions is now in focus of 
theoretical and experimental studies of relativistic heavy-ion collisions. 
This problem is one of the main 
motivations for the currently running beam-energy scan \cite{RHIC-scan} at the 
Relativistic Heavy-Ion Collider (RHIC) at Brookhaven National Laboratory (BNL) and 
low-energy-scan program \cite{SPS-scan} at Super Proton Synchrotron (SPS)
of the European Organization for Nuclear Research (CERN), as well as newly constructed 
Facility for Antiproton and Ion Research (FAIR) in Darmstadt \cite{FAIR} and the
Nuclotron-based Ion Collider Facility (NICA) in Dubna \cite{NICA}. 
In this paper I would like to argue that the baryon stopping 
in nuclear collision can be a sensitive probe of deconfinement onset. 

In fact, an  irregularity 
in the incident-energy dependence of the baryon stopping
 is very natural if the system 
undergoes a phase transition. Let us start with discussion of the conventional 
(i.e. one-fluid) hydrodynamics, which is applied to the whole process of 
the nuclear collision, e.g. from its compression stage
to the expansion stage up to freeze-out, like it is done 
in Refs.  \cite{Merdeev:2011bz,Mishustin:2010sd}.

The form of the resulting rapidity distribution of net-baryons 
depends on the spatial form of the produced fireball. 
If the fireball is almost spherical, the expansion of the fireball is essentially 
3-dimensional which results in a peak at the midrapidity in the rapidity distribution. 
This statement is a theorem that can be proved in few lines. 
If a the fireball is strongly deformed (compressed) in the beam direction, 
i.e. has a form of a disk, its expansion is approximately 1-dimensional that 
produces a dip at the midrapidity, which is confirmed by numerous simulations, 
see e.g. Ref. \cite{71}. In terms of the fluid mechanics this is a consequence 
of interaction of two rarefaction waves propagating from opposite peripheral 
sides of decaying disk toward its center  \cite{Land-Lif}. 
This speculation is very similar to that related to the elliptic flow: 
a strong elliptic flow results from a strongly deformed almond-shaped 
initial fireball with the  deformation of the resulting momentum 
distribution of particles being inverse to the  spatial deformation of the 
initial fireball. The physical mechanism here is precisely the same, only the 
expansion is developed in the transverse direction.

The next question is how this fireball is formed. This is already a matter of 
dynamics in the early compression stage of the nuclear collision. A softest point 
\cite{Hung:1994eq} characteristic of EoS's with a phase transition  
plays an important role in this compression dynamics. 
{The softest point in the equation of state is defined by a
minimum in the ratio of the pressure to 
the energy density at constant specific entropy (i.e. the entropy per baryon): $(P/\varepsilon)_{\sigma}$. 
The constancy of $\sigma$ is required because  it is conserved in the ideal 
hydrodynamics. At the softest point the system exhibits the weakest resistance 
to its compression as compared with that in adjacent regions of the EoS. 
In Ref. \cite{Hung:1994eq} and subsequent works attention was 
focused on the softest-point effect on the expansion stage of the collision. 
In particular, it
was argued \cite{Hung:1994eq} that at this point the comparatively small pressure prevents a fast expansion
and cooling of the system. Here I would like to argue that the same softest point also 
plays important role during the early compression stage, 
resulting in extra high energy and baryon densities of the produced fireball. 
}

{
Let us proceed in terms of the conventional (one-fluid) hydrodynamics.} 
At low energies the softest point is not reached in the collision process, 
the system remains  stiff and therefore the produced fireball is 
almost spherical. As a result, the baryon rapidity distribution is
peaked at the midrapidity. 
When the incident energy gets high enough, the softest-point region  
of the EoS starts to dominate during the compression stage, the system weaker  
resists to the compression and hence the resulting fireball becomes more 
deformed, i.e. more of the disk shape. Then its expansion is close to the 
1-dimensional pattern and, as a result, we have a dip at the midrapidity. 
With energy rise, the stiffness of the EoS (in the range relevant to compression stage) 
grows, the system stats to be more resistant to the compression and hence 
the produced fireball becomes less deformed. The expansion of this 
fireball results in a peak or, at least, to a weaker dip at the midrapidity 
as compared to that at the ``softest-point'' incident energy. 
With further energy rise, the initial kinetic pressure overcomes 
the stiffness of the EoS and makes the produced fireball strongly deformed 
again, which in its turn again results in a dip at the midrapidity.

Thus, even without any nonequilibrium, we can 
expect a kind of a ``peak-dip-peak-dip'' irregularity in the 
incident energy dependence of the form of the net-proton rapidity distributions.
Nonequilibrium also contributes to this irregularity. 
At a phase transformation\footnote{The term ``phase
  transition'' is deliberately avoided, since it usually implies
  thermal equilibrium.}  
 the hadronic degrees of freedom are changed to partonic ones. 
In particular, it means a change in cross sections which 
govern the nuclear stopping power. Therefore, this change also induces a
certain  irregularity in the stopping power.
Only the dip at the midrapidity in ultrarelativistic nuclear collisions 
has a different origin in the actual case of weak baryon stopping 
as compared with that in the conventional hydrodynamics. 
It occurs because the baryon charges of colliding nuclei traverse through 
each other.

It is important to emphasize that the ``peak-dip-peak-dip'' irregularity 
is a signal from the hot and dense stage of the nuclear collision.

In the present paper this qualitative pattern is illustrated by calculations 
within a model of the three-fluid 
dynamics (3FD) \cite{3FD} employing three different equations of state (EoS): a purely hadronic EoS   
\cite{gasEOS} (hadr. EoS), which was used in the major part of the 3FD simulations so far 
\cite{3FD,3FD-GSI07,3FDflow,3FDpt,3FDv2}, and two versions of EoS involving  
{  
deconfinement  
}
\cite{Toneev06}. 
These two versions are an EoS with the first-order phase transition (2-phase EoS) 
and that with a smooth crossover transition (crossover EoS). The softest points in these EoS's 
 are illustrated in Ref. 
\cite{Nikonov:1998dg}. The hadronic EOS \cite{gasEOS} possesses no softest point - 
stiffness of the EoS changes monotonously. 
Preliminary results of simulations with 
{  
deconfinement transitions 
}
have been already reported in Refs. \cite{Ivanov:2010cu,Ivanov:2011cb}. 
{
There the friction forces for the 2-phase and crossover scenarios were poorly tuned and hence 
the corresponding simulations poorly reproduced available experimental data. 
Therefore, the conjecture 
on an irregular behavior of the net-proton baryon  
was based on a certain trend of the results of simulations. 
}
Here I present calculations with thoroughly tuned friction forces in the quark-gluon phase, 
which made it possible to reasonably 
{
(and often better than in the hadronic scenario)
}
reproduce a great number of observables in 
{
a wider (than before \cite{Ivanov:2010cu,Ivanov:2011cb}) 
}
incident energy range 2.7 GeV  $\le \sqrt{s_{NN}}\le$ 39 GeV in terms of the 
center-of-mass energy. Details of this calculations and results on  
a great number of  bulk observables (rapidity and transverse spectra, flow observables and multiplicities) 
for various species and a large number of incident energies, their comparison with available data
will be reported elsewhere. 
Here I would like to focus on rapidity distributions of net-protons in central collisions 
of heavy nuclei in the AGS-SPS energy range. 
The distribution of net-protons is the best probe of the nuclear stopping 
in the absence of data on net-baryons. 

Concerning the fit of the friction force, few comments are in order. 
The friction in the hadronic phase was estimated in Ref. \cite{Sat90}.
Within hadronic scenario (hadr. EoS) we had to enhance this friction in order 
to reproduce the baryon stopping at high energies \cite{3FD}. 
Though such a enhancement is admissible in view of uncertainties 
of the estimated friction, the value of the enhancement looks too high. 
Indeed, at $\sqrt{s_{NN}}= 17.3$ GeV, i.e. at the top SPS energy, 
this enhancement exceeds the factor of 2. 
In scenarios with 
{  
deconfinement  
}
there is no need to modify the hadronic friction. 
This can be considered as a theoretical argument in favor of such scenarios.

\begin{figure}[bth]
\includegraphics[width=7.0cm]{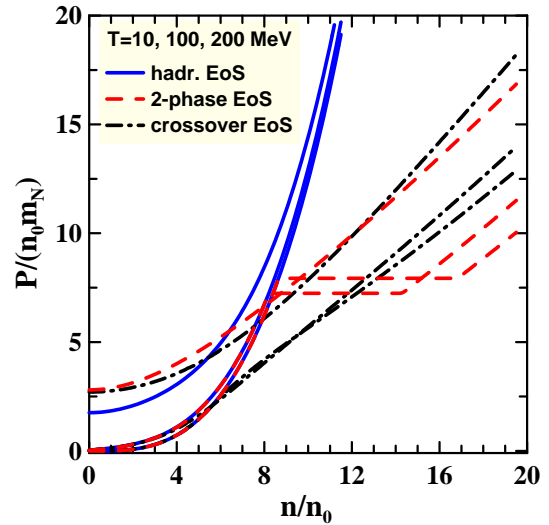}
 \caption{{
Pressure scaled by the product of normal nuclear density ($n_0=$ 0.15 fm$^{-3}$) and 
nucleon mass ($m_N$) versus baryon density scaled by the normal nuclear density
for three considered equations of state. Results are presented for three different
temperatures $T=$ 10, 100 and 200 MeV (bottom-up for corresponding curves).  
}} 
\label{fig3.1}
\end{figure}

\begin{figure*}[tbh]
\includegraphics[width=6.0cm]{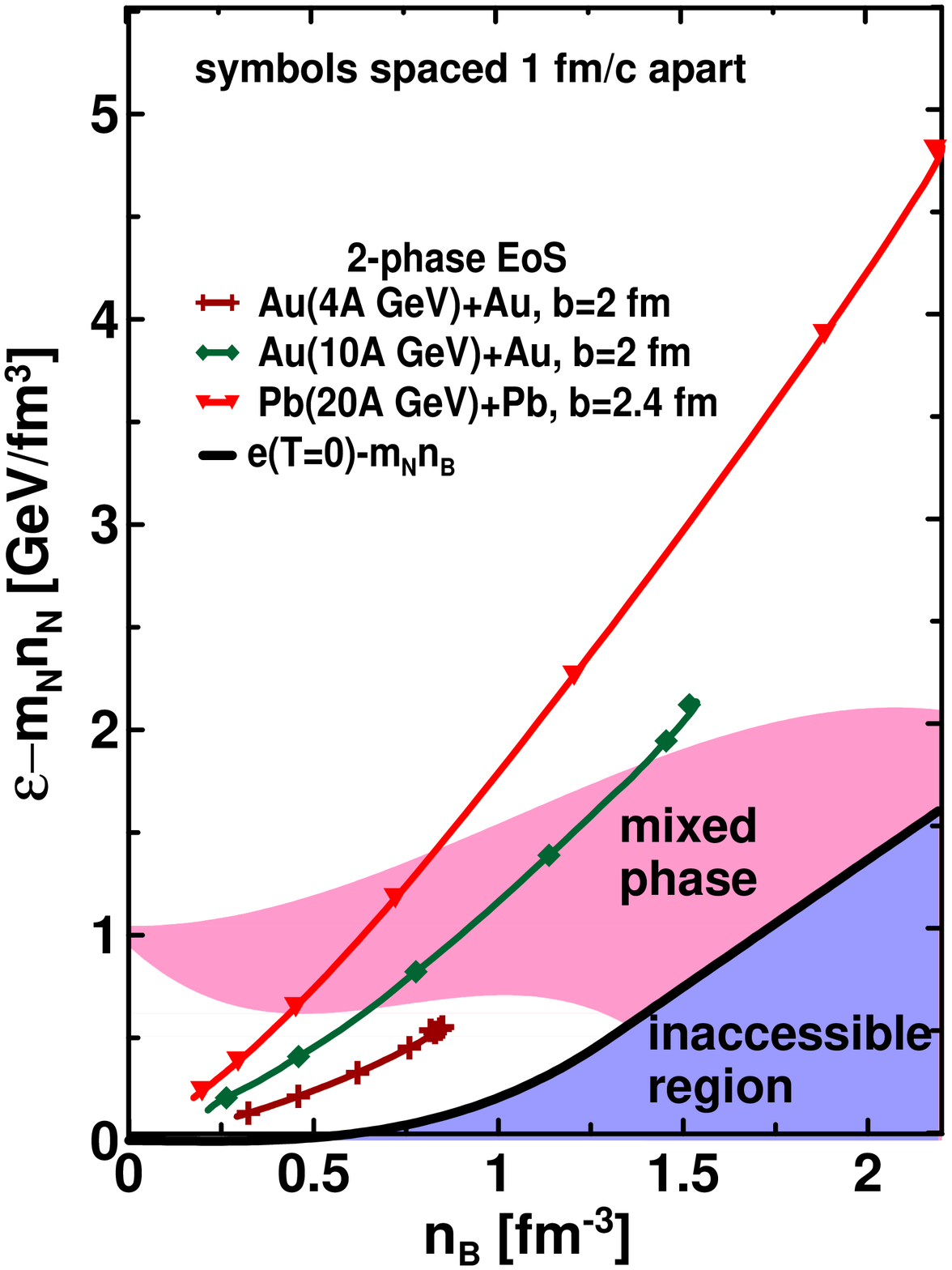}
\raisebox{9mm}{
\includegraphics[width=5.25cm]{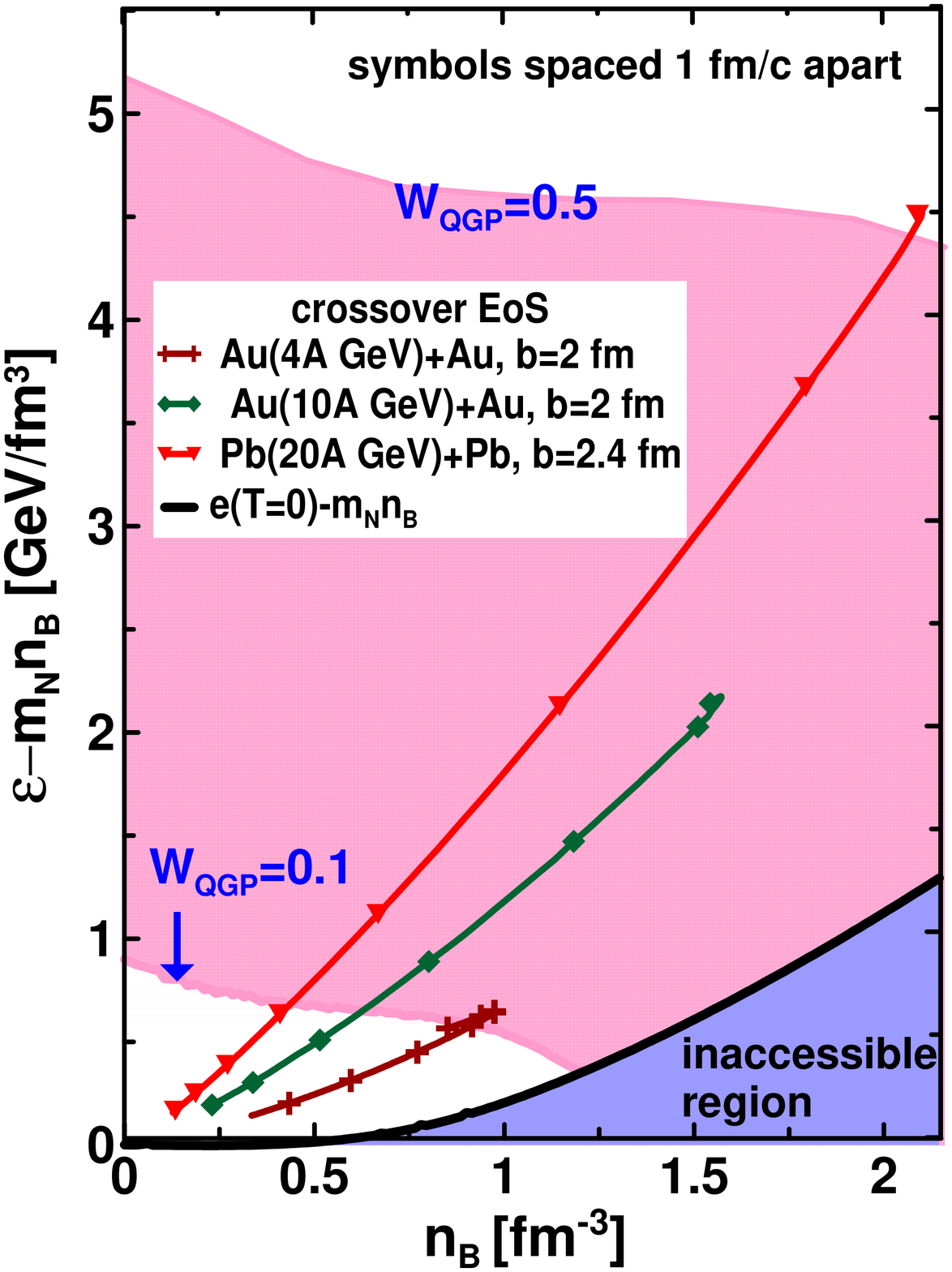}}
\vspace*{-10mm}
 \caption{
Dynamical trajectories of the matter in the central box of the
colliding nuclei  
(4fm$\times$4fm$\times \gamma_{cm}$4fm), where $\gamma_{cm}$ is the Lorentz
factor associated with the initial nuclear motion in the c.m. frame, 
for central ($b=$ 2 fm) collisions of Au+Au at 4$A$ and 10$A$ GeV 
energies and Pb+Pb at 20$A$ GeV. The trajectories are plotted in terms
of baryon density ($n_B$) and 
the energy density minus $n_B$ multiplied by the nucleon mass 
($\varepsilon - m_N n_B$). 
Only expansion stages of the
evolution are displayed.  The trajectories are presented 
for two EoS's: 2-phase EoS (left panel) and crossover EoS  (right panel).
Symbols on the trajectories indicate the time rate of the evolution:
time span between marks is 1 fm/c.  
For the 2-phase EoS (left panel) 
the shadowed ``mixed phase'' region is located between the borders, 
where the QGP phase start to raise ($W_{QGP}=$ 0) and becomes completely 
formed ($W_{QGP}=$ 1). 
For the crossover EoS  (right panel) the corresponding borders correspond to
values of the QGP fraction $W_{QGP}=$ 0.1 and $W_{QGP}=$ 0.5.
Inaccessible region is restricted by $\varepsilon(n_B,T=0)-m_N n_B$ from above. 
}
\label{fig3.2}
\end{figure*}

\section{Equations of State} 
\label{EOS}

{
Figure \ref{fig3.1} illustrates differences between three considered EoS's. 
The 
{  
deconfinement transition 
}
makes a EoS softer at high temperatures and/or densities.
The 2-phase EoS is based on the Gibbs construction, taking into account simultaneous conservation 
of baryon and strange charges. However, the displayed result looks very similar to 
the Maxwell construction, corresponding to conservation of only baryon charge, 
with the only difference that the plateau is slightly tilted, which is practically invisible. 
Application the Gibbs construction
in hydrodynamical simulations silently assumes that the inter-phase equilibration
in the mixed-phase region is faster than the hydrodynamical evolution. 
From the practical point of view, this assumption allows us to avoid problems
with instabilities in the spinodal region. However, from conceptual point of view
this assumption is not obvious. It would be better to consider a finite relaxation time 
for the inter-phase equilibration as it was proposed long ago \cite{Barz:1984az}. 
If this relaxation time is short enough, the unstable spinodal region will be still avoided, 
that is important for the numerical reasons. 
That is a long-term plan of refining the model. In the present calculations the 
instantaneous inter-phase equilibration is assumed. 
}

The 2-phase and crossover EoS's still differ even at very high densities. 
The latter means that the crossover transition constructed in Ref. \cite{Toneev06}
is very smooth. The hadronic fraction survives up to very high densities. In particular, 
this is seen from Fig. \ref{fig3.2}: the fraction of the quark-gluon phase 
($W_{QGP}$) reaches value of 0.5 only at very high energy densities. 
In this respect, this version of the crossover EoS certainly contradicts results of the 
lattice QCD calculations, where a fast crossover, at least at zero chemical potential, 
was found \cite{Aoki:2006we}. 
Therefore, a true EoS is somewhere in between  the crossover and 2-phase EoS's 
of Ref. \cite{Toneev06}.

Figure \ref{fig3.2} demonstrates that the onset of  
{  
deconfinement  
}
in the calculations happens at top-AGS--low-SPS energies. 
Similarly to Ref. \cite{Randrup07}, 
the figure displays dynamical 
trajectories of the matter in a central box placed around the
origin ${\bf r}=(0,0,0)$ in the frame of equal velocities of
colliding nuclei:  $|x|\leq$ 2 fm,  $|y|\leq$ 2 fm and $|z|\leq$
$\gamma_{cm}$ 2 fm, where $\gamma_{cm}$ is Lorentz
factor associated with the initial nuclear motion in the c.m. frame.  
Initially, the colliding nuclei are placed symmetrically with respect
to the origin ${\bf r}=(0,0,0)$, $z$ is the direction of the beam.
At a given density $n_B$, the zero-temperature
compressional energy, $\varepsilon(n_B,T=0)$, provides a lower bound on
the energy density $\varepsilon$, so the accessible region is correspondingly
limited. 
In the case of the crossover EoS only the region of the mixed phase between 
$W_{QGP}=$ 0.1 and $W_{QGP}=$ 0.5 is displayed, since in fact the mixed phase
occupies the whole ($\varepsilon$-$n_B$) region. 
The $\varepsilon$-$n_B$ representation
is chosen because these densities 
are dynamical quantities and, therefore, are suitable to compare
calculations with different EoS's.

\begin{figure*}[bt]
\hspace*{-14mm}\includegraphics[width=8.9cm]{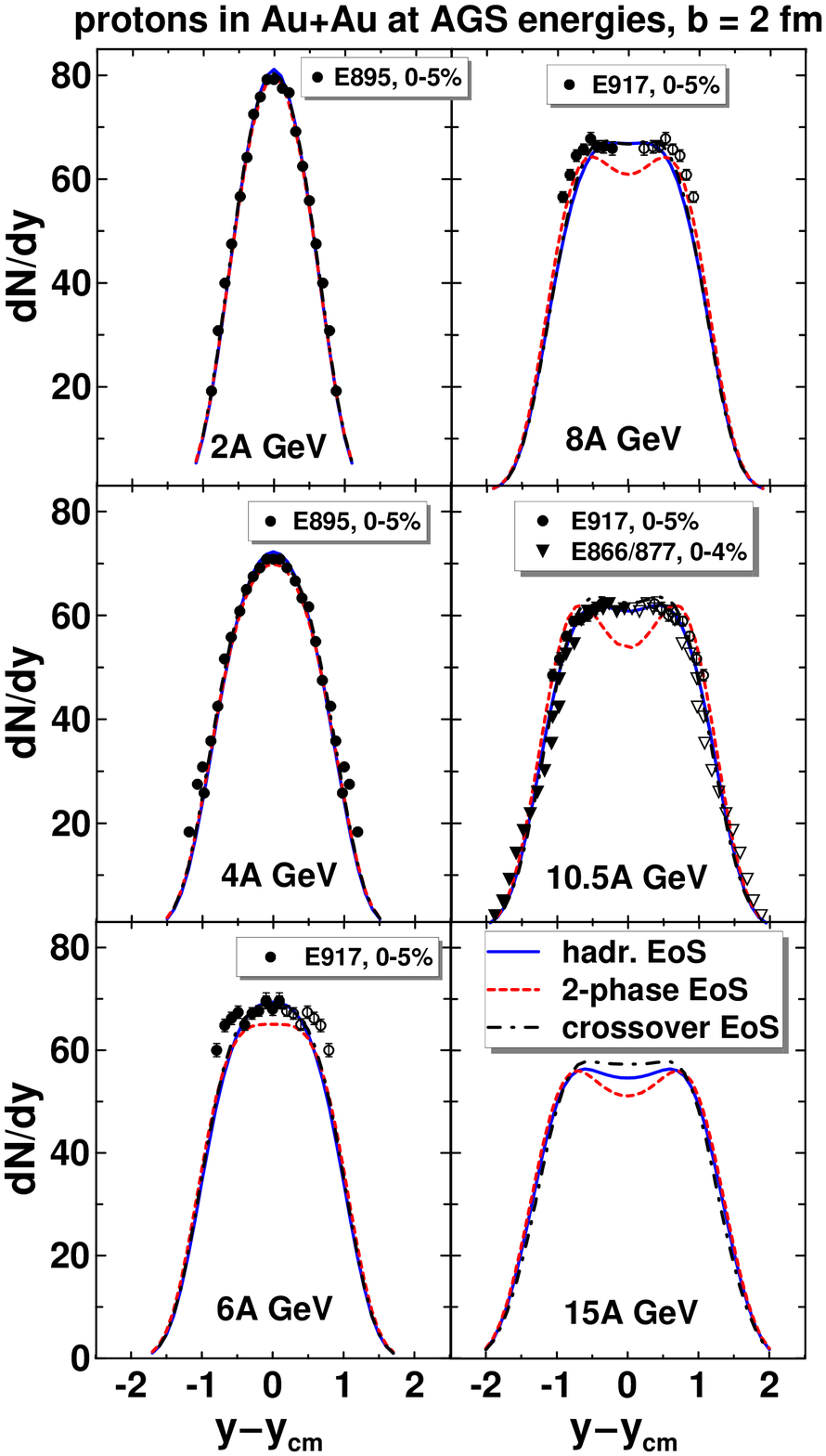}\hspace*{-10mm}
\includegraphics[width=8.9cm]{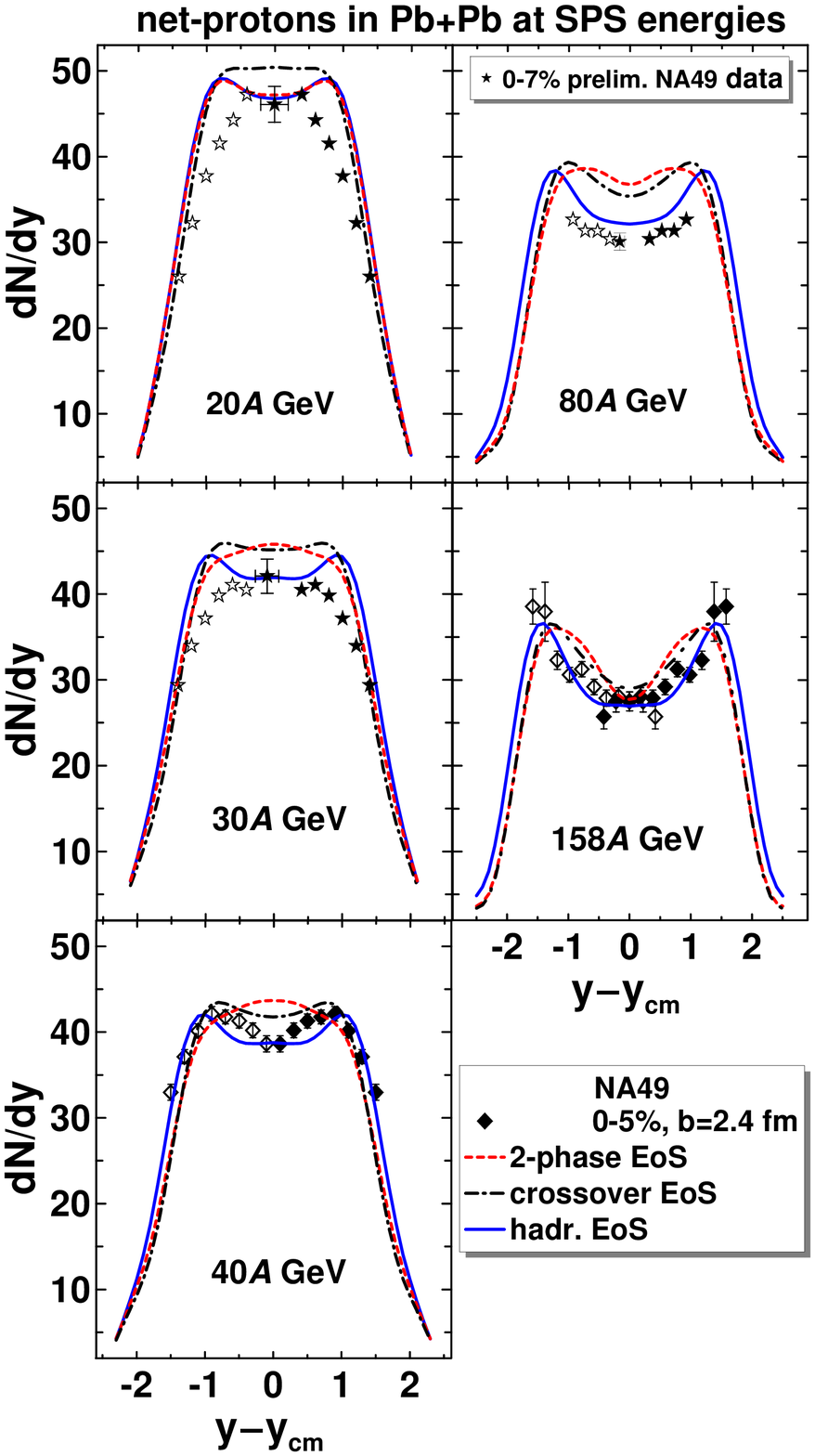}\hspace*{-14mm}
\vspace*{-10mm}
 \caption{
Rapidity spectra of protons (for AGS energies, see left panel)
and net-protons $(p-\bar p)$ (for SPS energies, see right panel) from  
central collisions of Au+Au (AGS) and Pb+Pb (SPS). 
Experimental data are from
collaborations E895 \cite{E895}, E877 \cite{E877},
E917 \cite{E917}, E866 \cite{E866}, and 
NA49 \cite{NA49-1,NA49-04,NA49-06,NA49-07,NA49-09}. 
The percentage shows the fraction of the total reaction cross section, 
corresponding to experimental selection of central events. 
Feedback of weak decays into $p$ and $\bar p$ yields is disregarded. 
} 
\label{fig4.1}
\end{figure*}

Only expansion stages of the evolution are displayed, where the matter
in the box is already thermally equilibrated.   
Evolution proceeds from the top point of the trajectory downwards.
Subtraction of the $m_N n_B$ term is taken for the sake of suitable 
representation of the plot. 
The size of the box was chosen 
to be large enough that the amount of matter in it can be
representative to conclude on the onset of 
{  
 deconfinement  
}
and to be small enough to consider the matter in it as a homogeneous
medium. Nevertheless, the matter in the box still amounts to a minor part
of the total matter of colliding nuclei.  
Therefore, only the minor part of the total matter undergoes  the
{  
deconfinement transition 
}
at 10$A$ GeV energy. 

As seen, the 
{  
deconfinement transition 
}
starts at the top AGS energies in both cases. 
It gets practically completed at low SPS energies in the case of the case of the   
2-phase EoS. In the crossover scenario it lasts till very high incident energies. 
The trajectories for two different  EoS's are nevertheless very similar at 
 displayed energies. Apparently, it happens because the friction in the quark-gluon phase 
 was selected in such a way that both scenarios reasonably reproduce  
 available data at high energies.

\section{Proton and Net-Proton rapidity distributions}
\label{rapidity distributions}

A direct measure of the baryon stopping is the
net-baryon (i.e. baryons-minus-antibaryons) rapidity distribution. However, since experimental
information on neutrons is unavailable, we have to rely on net-proton (i.e. proton-minus-antiproton) data. 
Presently there exist experimental data on proton (or net-proton) rapidity spectra at 
AGS \cite{E895,E877,E917,E866} and 
SPS \cite{NA49-1,NA49-04,NA49-06,NA49-07} energies. 
These data were analyzed within various models  
\cite{3FD,3FD-GSI07,Bratk09,Merdeev:2011bz,Mishustin:2010sd,Bleicher09,Bratk04,WBCS03,Bratk02,Larionov07,Larionov05}.

Figure \ref{fig4.1} presents calculated rapidity distributions of protons (for AGS energies) 
and net-protons (for SPS energies) and their comparison with available data. Notice that 
difference between protons and net-protons, as well as a contribution of weak decays
to these yilds are negligible at the AGS energies,  
see compilation of experimental data in Ref. \cite{Andronic:2005yp}. 
Contribution of weak decays of strange hyperons into proton yield was disregarded
in accordance with measurement conditions of the NA49 collaboration. 
Correspondence between the fraction of the total cross section related to a data set 
and a mean value of  the impact parameter was read off from the paper \cite{Alt:2003ab}
in case of NA49 data. 
For Au+Au collisions it was approximately estimated proceeding from geometrical considerations.

As seen from Fig. \ref{fig4.1}, at lower AGS energies all EoS's predict the same results, 
since at these energies only hadronic parts of all EoS's are relevant, see Fig. \ref{fig3.2}.   
Results of the 
2-phase EoS start to differ from those of the hadr. and crossover EoS's beginning
from 6$A$ GeV:   
the 2-ph.-EoS distributions reveal a dip at the midrapidity. 
This dip contradicts the available experimental data and is very robust: 
variation of the friction in a 
wide range does not remove this dip. Therefore, it is a direct consequence 
of the onset of the first-order phase transition, which starts precisely at these 
energies in the 2-ph.-EoS scenario, see Fig. \ref{fig3.2}. 
This dip survives 
even in one-fluid calculations \cite{Merdeev:2011bz,Mishustin:2010sd} 
involving the 1st-order phase transition in spite of 
immediate baryon stopping inherent in the one-fluid model. 
In 3FD calculations,  this dip changes into midrapidity peak at higher energies (30$A$ GeV and  40$A$ GeV). 
With further energy rise ($E_{lab}>$ 40$A$ GeV) the midrapidity peak again transforms into a dip, 
see also  Fig. \ref{fig4.1}. 
The latter dip is already a normal behavior which takes place at arbitrary high energies, 
{
which is associated with growing transparency of the colliding nuclei rather than with  
production of strongly deformed fireball of completely stopped matter, as it was discussed 
at the example of the conventional (one-fluid) hydrodynamics. 
Thus, the ``peak-dip-peak-dip'' irregularity indeed results from essential 
interplay between softening of the EoS and incomplete stopping of the colliding matter. 
}


The experimental distributions exhibit a qualitatively similar
behavior as that in the 2-phase-EoS scenario: after a plateau or a shallow midrapidity dip   
at the energy of 8$A$ and 10$A$ GeV, a well pronounced peak at 20$A$ GeV is again observed. 
However, quantitatively the 
 2-phase-EoS results certainly disagree with data at 8$A$ GeV, 10$A$ GeV and  40$A$ GeV energies. 
They also disagree with data 20$A$ GeV and  30$A$ GeV, which  
however have preliminary status, and hence it is too early to draw any conclusions 
from comparison with them.  The 2-phase-EoS behavior is in contrast with that for the hadronic-EoS scenario, 
where the form of distribution in central collisions gradually evolve from peak at the 
midrapidity to a dip. The case of the crossover EoS is intermediate: only
 a shallow dip occurs at 10$A$ and 15$A$ GeV  while at 
20$A$ GeV the distribution looks like a plateau.

Predictions of different scenarios diverge to the largest extent
in the energy region  8$A$  GeV  $\leq E_{lab} \leq$ 40$A$  GeV. 
Unfortunately data at 20$A$ and 30$A$  GeV still have a preliminary status and 
disagree with any considered scenario. 
Updated experimental results at energies 20$A$ and 30$A$  GeV are badly needed 
to pin down the preferable EoS and to check the hint to the zigzag behavior of the 
type ``peak-dip-peak-dip'' in the net-proton rapidity distributions.

\section{Analysis of ``peak-dip-peak-dip'' irregularity}

\begin{figure*}[tb]
\vspace*{-54mm}
\includegraphics[width=13.1cm]{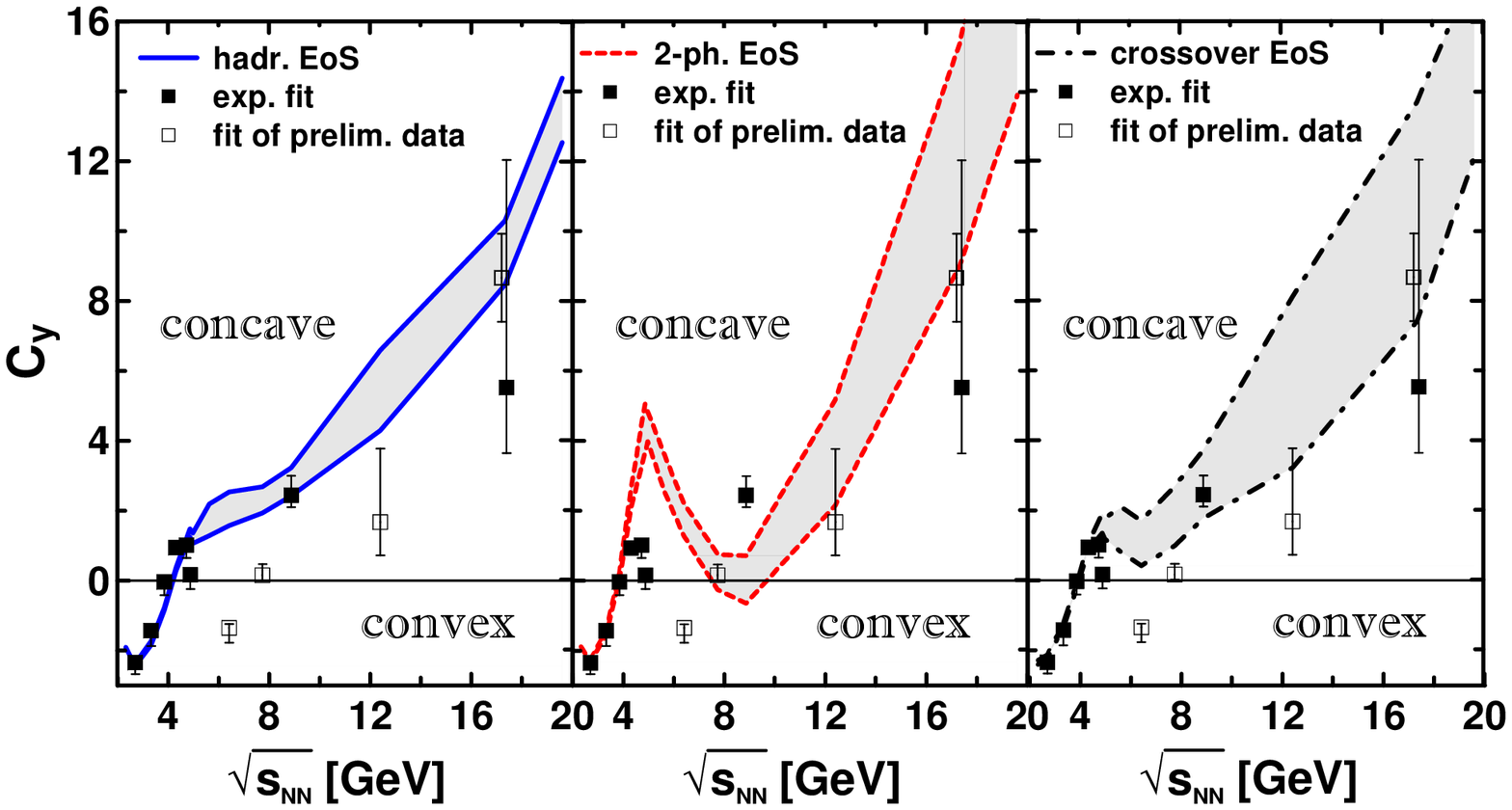}
\vspace*{-59mm}
 \caption{
Midrapidity  reduced curvature  
   [see. Eq. (\ref{Cy})] 
of the (net)proton rapidity
   spectrum as a function of the center-of-mass energy
 of colliding nuclei as deduced from experimental data and predicted
 by 3FD calculations with  different EoS's: the hadronic EoS
   (hadr. EoS) \cite{gasEOS} (left panel),  the EoS involving a first-order phase
 transition (2-ph. EoS, middle panel) and the EoS with a crossover transition (crossover EoS, right panel) 
 into the quark-gluon phase \cite{Toneev06}. 
Upper bounds of the shaded areas correspond to fits confined in the region of $|y-y_{cm}|/y_{cm}<0.7$,  
lower bounds -- $|y-y_{cm}|/y_{cm}<0.5$. 
}  
\label{fig4a}
\end{figure*}
\begin{figure}[htb]
\vspace*{-4mm}
\includegraphics[width=5.35cm]{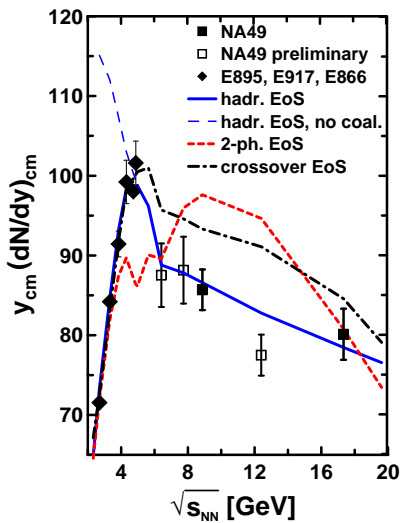}
\vspace*{-6mm}
 \caption{
The same as in Fig. \ref{fig4a} but for 
midrapidity  value 
of the (net)proton rapidity
   spectrum scaled by $y_{cm}$. 
The thin long-dashed line corresponds to the 
hadr.-EoS  calculation 
without fragment  production, i.e. without coalescence. 
\vspace*{-3mm}
}  
\label{fig4b}
\end{figure}

\begin{figure*}[tbh]
\includegraphics[width=5.35cm]{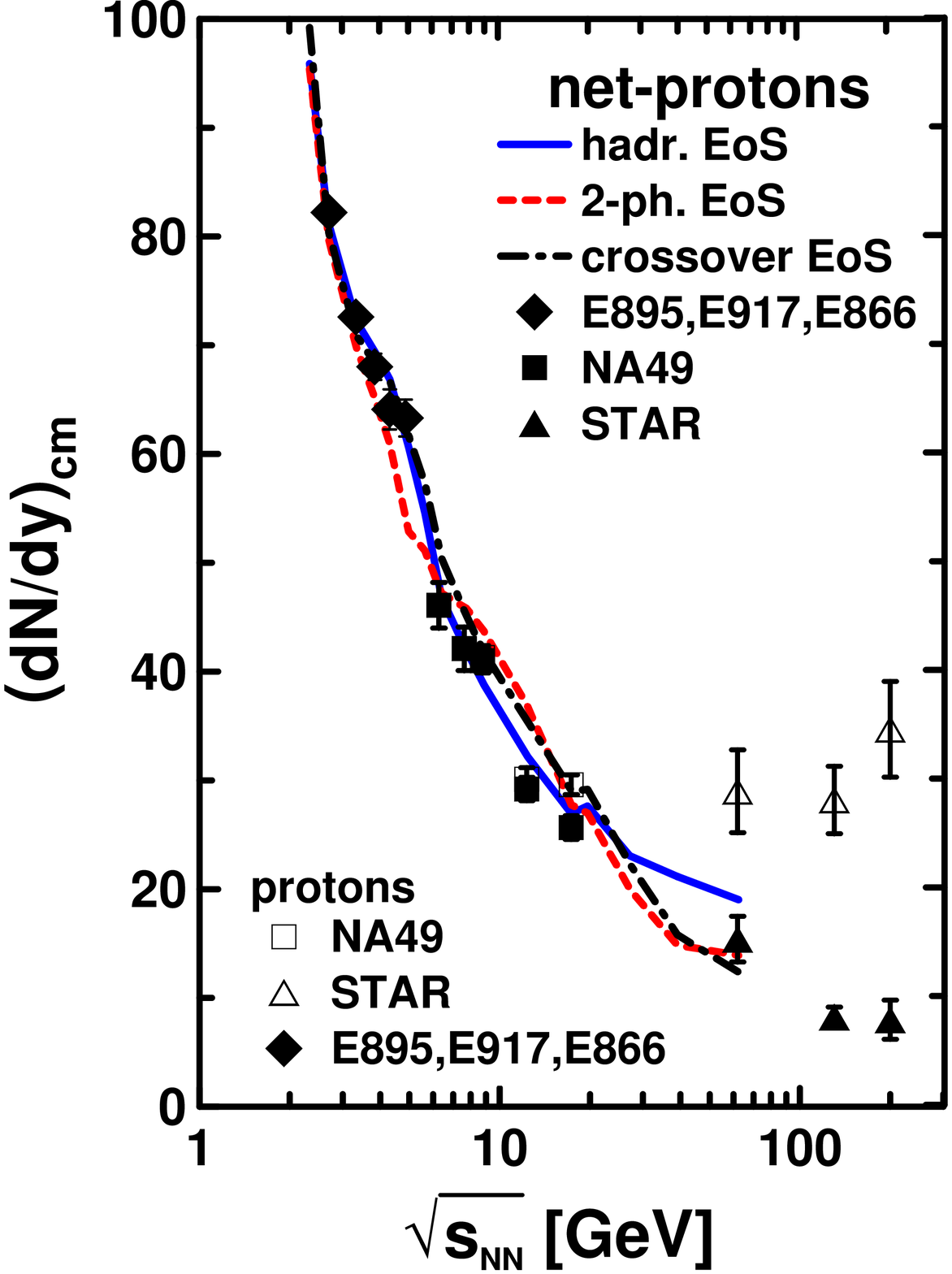}
\includegraphics[width=5.35cm]{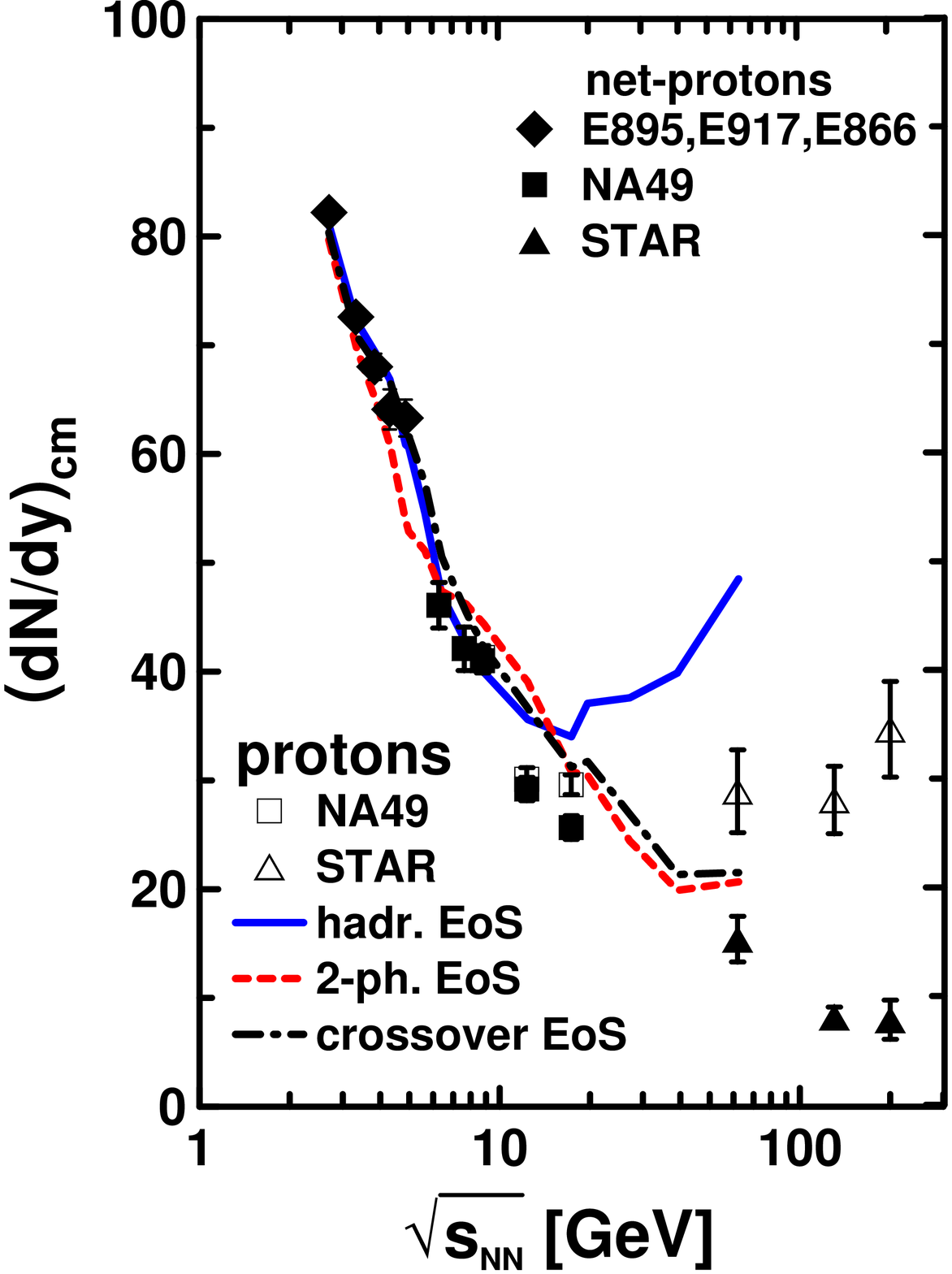}
 \caption{
The same as in Fig. \ref{fig4b} but for 
midrapidity  values 
of the net-proton (left panel) and proton (right panel) rapidity
   spectrum in conventional representation, i.e. without scaling. 
%
Experimental data are from
collaborations E895 \cite{E895}, E877 \cite{E877},
E917 \cite{E917}, E866 \cite{E866},  
NA49 \cite{NA49-1,NA49-04,NA49-06,NA49-07,NA49-09} and 
STAR  \cite{STAR09}. 
}  
\label{fig4c}
\end{figure*}
In order to quantify the above-discussed 
``peak-dip-peak-dip'' irregularity, it is useful to make use of the method 
proposed in Ref. \cite{Ivanov:2010cu}. For this purpose
the data on net-proton rapidity distributions are fitted  by a simple formula 
\begin{eqnarray}
\label{2-sources-fit} 
\frac{dN}{dy}&=&  
a \left(
\exp\left\{ -(1/w_s)  \cosh(y-y_{cm}-y_s) \right\}
\right.
\cr
&+&
\left.
\exp\left\{-(1/w_s)  \cosh(y-y_{cm}+y_s)\right\}
\right)
\end{eqnarray}
where $a$, $y_s$ and $w_s$ are parameters of the fit. The form
(\ref{2-sources-fit}) is a sum of two thermal sources shifted by $\pm
y_s$ from the midrapidity. The width $w_s$ of the sources can be
interpreted as $w_s=$ (temperature)/(transverse mass), if we assume
that collective velocities in the sources have no spread with respect
to the  source rapidities $\pm y_s$. The parameters of the two sources
are identical (up to the sign of  $y_s$) because  only 
collisions of identical nuclei are considered.

The above fit has been done by the least-squares method and 
applied to both available data and results of calculations.  
The fit was performed in the rapidity range $|y-y_{cm}|/y_{cm}<0.7$. 
The choice of this range is dictated by the data. As a rule, the data
are available in this rapidity range, sometimes the data range is even
more narrow (80$A$ GeV and new data at 158$A$ GeV \cite{NA49-09}). 
I put the above restriction in order to treat different data in
approximately the same rapidity range. 
Another reason for this cut is 
that the rapidity range should not be too wide in order to
exclude contribution of cold spectators.
The fit in the rapidity range $|y-y_{cm}|/y_{cm}<0.5$ has been also done 
in order to estimate uncertainty of the fit parameters associated with the 
choice of fit range.

A useful quantity, which characterizes the shape of the rapidity distribution, is
 a reduced curvature of the spectrum at the
midrapidity defined as follows 
\begin{eqnarray}
\label{Cy} 
C_y &=& 
\left(y_{cm}^3\frac{d^3N}{dy^3}\right)_{y=y_{cm}}
\big/ \left(y_{cm}\frac{dN}{dy}\right)_{y=y_{cm}}
\cr
&=&  
(y_{cm}/w_s)^2 \left(
\sinh^2 y_s -w_s \cosh y_s 
\right). 
\end{eqnarray}
The factor $1/\left(y_{cm}dN/dy\right)_{y=y_{cm}}$
is introduced in order to get rid of overall normalization of the
spectrum. The second part of Eq. (\ref{Cy}) presents 
this curvature in terms of parameters of fit (\ref{2-sources-fit}).
The reduced curvature, $C_y$, and the midrapidity value, 
$\left(y_{cm}dN/dy\right)_{y=y_{cm}}$, are two independent quantities quantifying the 
the spectrum in the midrapidity range. Excitation functions of these quantities
deduced both from 
experimental data and from 
results of the 3FD calculations 
with different EoS's are displayed in Figs. \ref{fig4a} and \ref{fig4b}. 
Notice that a maximum in 
$y_{cm} (dN/dy)_{cm}$ at $\sqrt{s_{NN}}=4.7$ GeV happens only because the light
fragment production becomes negligible above this energy. The 3FD
calculation without coalescence (i.e. without the light fragment production)
reveals a monotonous decrease of $y_{cm} (dN/dy)_{cm}$ beginning from 
the lowest energy considered here.

To evaluate errors of $C_y$ values deduced from data, I estimated the errors
produced by the least-squares method, as well as performed fits in
different the rapidity ranges: $|y-y_{cm}|/y_{cm}<0.5$ and 
$|y-y_{cm}|/y_{cm}<0.7$, where it is appropriate. 
Problems were met in fitting the data at 
80$A$  GeV \cite{NA49-07} and the  new data at 158$A$ GeV \cite{NA49-09}. 
These data do not go beyond the side maxima in the rapidity
distributions. This results in large uncertainty in the parameters.  
In particular, because of this problem I 
keep the old data at 158$A$ GeV 
\cite{NA49-1} in the analysis. 
The error bars present largest uncertainties among mentioned above. 
The upper errors of  158$A$-GeV 80$A$-GeV points
results from the uncertainty of the narrow rapidity range. 
The uncertainty associated with the choice of the 
rapidity range  
turned out to be a dominant one 
in the case computed data. 
Therefore, in Fig.  \ref{fig4a} results for the 
computed spectra are presented by shaded areas with borders corresponding 
to the fit ranges $|y-y_{cm}|/y_{cm}<0.7$ and $|y-y_{cm}|/y_{cm}<0.5$. 
In Fig. \ref{fig4b} the midrapidity  values 
of the rapidity spectra were taken directly from experimental data and calculated results. 
Therefore, only experimental error bars are displayed there.

Since experimental data at AGS and RHIC energies were taken 
from Au+Au collisions  
while at SPS the Pb+Pb collisions were studied, the calculations were performed 
respectively for Au+Au ($b=$ 2 fm) and Pb+Pb ($b=$ 2.4 fm) central collisions. 
In fact, at the same incident energy the computed results for 
Pb+Pb collisions at $b=2.4$ fm are very close to those for Au+Au at $b=2$ fm.
Therefore, the corresponding irregularity of the energy dependence of the fit parameters
is negligible.

The irregularity in data  is distinctly seen here 
as a zigzag irregularity in the energy dependence of $C_y$. 
Of course, this is only a hint to irregularity since this zigzag is formed only due to 
preliminary data of the NA49 collaboration. 
A remarkable observation is that 
the $C_y$ excitation function in the first-order-transition
scenario manifests qualitatively the same zigzag irregularity 
(left panel of Fig. \ref{fig4a}) as
that in the data fit, while the hadronic  scenario produces purely monotonous 
behavior. 
The crossover EoS represents a very smooth transition, as mentioned above. Therefore, 
it is not surprising that it produces only a weak wiggle in $C_y$.

This zigzag irregularity of the first-order-transition scenario is also 
reflected in the midrapidity values of the (net)proton rapidity
spectrum (Fig. \ref{fig4b}). 
In the conventional representation of the data 
without multiplying by $y_{cm}$, the irregularity of the
$(dN/dy)_{cm}$ data is hardly visible (Fig. \ref{fig4c}). 
Moreover, the difference between predictions resulted from different EoS's is also scarcely discernible. 
Thus, the scaled representation of Fig. \ref{fig4b} serves as a zoom revealing differences. 
In Fig. \ref{fig4c} not only net-protons but also proton midrapidity  values 
are displayed in a wider energy range. However, results for top energy  $\sqrt{s_{NN}}=$ 62.4 GeV 
are very approximate, 
since more accurate computation requires unreasonably high memory and CPU time. 
As seen, a visible difference between net-protons and proton data, 
as well between predictions of hadronic EoS and EoS's with 
{  
deconfinement  
}
starts only at RHIC energies. 
Similar (but based on 
different, double-gaussian fit) analysis was performed in Ref. \cite{MehtarTani:2011uq}. 
It found no irregularities in excitation functions of the corresponding parameters, in particular,  
because the  analysis of Ref. \cite{MehtarTani:2011uq} 
was restricted to energies $E_{lab} \geq$ 20$A$  GeV.

As it was pointed out in the introduction, 
 the ``peak-dip-peak-dip'' irregularity is very natural in a system 
undergoing a 
{  
phase or crossover transition. 
}
First, it is associated with the softest point 
of a EoS. Therefore, it is not surprising that the irregularity is weaker 
in the crossover scenario than in the first-order-transition one. Indeed, 
the softest points in the crossover EoS is less 
pronounced than in the first-order-transition one \cite{Nikonov:1998dg}. 
There is no softest point in the hadronic EoS and 
hence there is no irregularity.

The second reason of this irregularity is a change in the nonequilibrium regime. 
The 3FD model takes into account the leading nonequilibrium of the nuclear 
collision associated with a finite stopping power of the nuclear matter. 
It simulates the 
finite stopping power by means of friction between three fluids. Naturally, 
this friction changes when  
{  
deconfinement  
}
happens. 
In the case of the crossover scenario this change in the friction is very 
smooth. Therefore, it does not contribute to the irregularity. 
At the same time this change in the friction enhances the irregularity in 
the first-order-transition scenario. 
As it was demonstrated in 
Ref. \cite{Ivanov:2010cu}, if the same friction is used in both phases, the
reduced curvature calculated with the 2-phase EoS reveals a wiggle
 behavior in $C_y$ but with considerably smaller amplitude  
as compared with zigzag in actual calculations with different frictions 
in different phases. These different frictions appear quite naturally 
in the 3FD model. The hadronic friction was estimated in Ref. \cite{Sat90} and works 
well at lower AGS energies. Therefore, there are no reasons to modify it. 
The partonic friction, while not microscopically estimated, is fitted to reproduce 
data at high incident energies. This is a reason to believe that it is a proper 
choice.

\section{Conclusion}

An irregularity in the baryon stopping is a natural consequence of  
{  
 deconfinement  
}
occurring in the compression stage of a nuclear collision. It is a combined effect 
of the softest point of a EoS and 
a change in the nonequilibrium regime from hadronic to partonic one.  
It is important to emphasize that this irregularity 
is a signal from the hot and dense stage of the nuclear collision.

{
Of course, this irregularity does not directly indicate deconfinement, i.e. transition to 
quark-gluon degrees of freedom. It only indicates softening of the EoS (occurence 
of the softest point) in a certain domain of 
the phase space. In principle, this softening may happen due to other reasons rather than 
onset of deconfinement. Nevertheless, I reason in terms of the deconfinement because 
this softening is associated with its onset in the considered EoS's. 
}

In order to illustrate this irregularity, calculations within the 3FD model were performed 
with three different equations of state, i.e. with a purely hadronic EoS  
\cite{gasEOS},  and also with two versions of EoS involving  
{  
deconfinement  
}
\cite{Toneev06}: an EoS with the first-order phase transition  
and that with a smooth crossover transition.
The crossover transition constructed in Ref. \cite{Toneev06}
is very smooth. The hadronic fraction survives up to very high energy densities. 
In this respect, this version of the crossover EoS certainly contradicts results of the 
lattice QCD calculations, where a fast crossover, at least at zero chemical potential, 
was found \cite{Aoki:2006we}. 
Therefore, a true EoS is somewhere in between  the crossover and 2-phase EoS's 
of Ref. \cite{Toneev06}.

It is found that predictions within the  first-order-transition scenario
reveal a ``peak-dip-peak-dip'' irregularity in the 
incident energy dependence of the form of the net-proton rapidity distributions in central collisions. 
At low energies, rapidity distributions 
have a peak at the midrapidity. With the incident energy rise it transforms into 
a dip, then again into a peak, and with further energy rise 
the midrapidity peak again change into a dip, which already survives up to arbitrary high energies.
The behavior the type of ``peak-dip-peak-dip'' in central collisions within
the 2-phase-EoS scenario is very robust 
with respect to variation of the model parameters in a 
wide range. 
This behavior is in contrast with that for the hadronic-EoS scenario, 
where the distribution form gradually evolve from peak at the 
midrapidity to a dip. The case of the crossover EoS is intermediate. Only  
a weak wiggle of the type of ``peak-dip-peak-dip'' takes place. 

Experimental data also revel a trend of the ``peak-dip-peak-dip'' irregularity 
in the energy range 8$A$ GeV $\le E_{lab}\le$ 40$A$ GeV, which qualitatively similar 
to that in the first-order-transition scenario while quantitatively differ. 
However, the  this experimental trend is based on 
preliminary data at energies of 20$A$ GeV and  30$A$ GeV. 
In general, predictions of different scenarios diverge to the largest extent
in the energy region  8$A$  GeV  $\leq E_{lab} \leq$ 40$A$  GeV. 
Therefore, updated experimental results at energies 20$A$ and 30$A$  GeV are badly needed 
to pin down the preferable EoS and to check the hint to the zigzag behavior of the 
type ``peak-dip-peak-dip'' in net-proton rapidity distributions. 
Moreover, 
it would be highly desirable if new data in this energy range are taken 
within the same experimental setup and at the same centrality selection.  
Hopefully 
such data will come from new accelerators FAIR at GSI and NICA at
Dubna.

In order to quantify the observed  
``peak-dip-peak-dip'' irregularity, the analysis of the distribution shape 
proposed in Refs. \cite{Ivanov:2010cu,Ivanov:2011cb} was applied. 
This method is based on calculation of the reduced curvature of the spectrum at the
midrapidity $C_y$. In tems of $C_y$
the irregularity in data is distinctly seen  
as a zigzag irregularity in the energy dependence of $C_y$. 
The energy location of this zigzag anomaly
coincides with the previously observed  
anomalies for other hadron-production properties at the
  low SPS energies \cite{Alt:2007fe,Gazdzicki:1998vd}. 
A remarkable observation is that 
the $C_y$ energy dependence in the first-order-transition
scenario manifests qualitatively the same zigzag irregularity, as
that in the data fit, though quantitatively does not reproduce the data fit. 
The hadronic  scenario produces purely monotonous 
behaviour. 
The crossover EoS represents a very smooth transition, as mentioned above. Therefore, 
it is not surprising that it results in only a weak wiggle in $C_y$.

Here the distribution of net-baryons was discussed because net-baryons have advantage 
of being confined by the baryon number conservation. In general, 
an analysis of the form of  particle spectra looks very promising. 
For instance, in Ref. \cite{Steinheimer:2012bp} it was proposed 
to analyze the incident-energy dependence of the width of 
pion rapidity distributions.  This width can be associated 
with the sound velocity in the dense stage of the reactions. 
It is found
that the sound velocity has a local minimum (indicating a softest point in the EoS)
in the range of $\sqrt{s_{NN}}=$ 4-9 GeV, which coincides with the range of 
the ``peak-dip-peak-dip'' irregularity discussed in this paper and that of 
the previously observed  
anomalies for other hadron-production properties \cite{Alt:2007fe,Gazdzicki:1998vd}.


Fruitful discussions with A.V.~Merdeev,
I.N. Mishustin, L.M. Satarov 
and D.N. Voskresensky
are gratefully acknowledged. 
I am grateful to A.S. Khvorostukhin, V.V. Skokov,  and V.D. Toneev for providing 
me with the tabulated 2-phase and crossover EoS's. 
The calculations were performed at the computer cluster of GSI (Darmstadt). 
This work was supported by The Foundation for Internet Development (Moscow)
and also partially supported  by  
the Russian Ministry of Science and Education 
grant NS-215.2012.2.

\end{document}